Best practices for estimating and reporting epidemiological delay distributions of infectious diseases using public health surveillance and healthcare data

Kelly Charniga[1], Sang Woo Park[2], Andrei R. Akhmetzhanov[3], Anne Cori[4], Jonathan Dushoff[5,6,7], Sebastian Funk[8,9], Katelyn M. Gostic[10], Natalie M. Linton[11], Adrian Lison[12], Christopher E. Overton[13,14,15], Juliet R.C. Pulliam[10], Thomas Ward[14], Simon Cauchemez[1], Sam Abbott[8,9]

[1] Mathematical Modelling of Infectious Diseases Unit, Institut Pasteur, Université Paris Cité, CNRS UMR 2000, Paris, France
[2] Department of Ecology and Evolutionary Biology, Princeton University, Princeton, New Jersey, USA
[3] College of Public Health, National Taiwan University, Taipei, Taiwan
[4] MRC Centre for Global Infectious Disease Analysis, School of Public Health, Imperial College London, London, UK
[5] Departments of Mathematics & Statistics and Biology, McMaster University, Hamilton, Ontario, Canada
[6] Department of Biology, McMaster University, Hamilton, Ontario, Canada
[7] M. G. DeGroote Institute for Infectious Disease Research, McMaster University, Hamilton, Ontario, Canada
[8] Department of Infectious Disease Epidemiology and Dynamics, London School of Hygiene & Tropical Medicine, London, UK
[9] Centre for Mathematical Modelling of Infectious Diseases, London School of Hygiene & Tropical Medicine, London, UK
[10] Center for Forecasting and Outbreak Analytics, U.S. Centers for Disease Control and Prevention, Atlanta, Georgia, USA
[11] Graduate School of Medicine, Hokkaido University, Sapporo-shi, Hokkaido, Japan
[12] Department of Biosystems Science and Engineering, ETH Zurich, Zurich, Switzerland
[13] Department of Mathematical Sciences, University of Liverpool, Liverpool, UK
[14] All Hazards Intelligence, Infectious Disease Modelling Team, Data Analytics and Surveillance, UK Health Security Agency, UK
[15] Department of Mathematics, University of Manchester, Manchester, UK

**Abstract**

Epidemiological delays, such as incubation periods, serial intervals, and hospital lengths of stay, are among key quantities in infectious disease epidemiology that inform public health policy and clinical practice. This information is used to inform mathematical and statistical models, which in turn can inform control strategies. There are three main challenges that make delay distributions difficult to estimate. First, the data are commonly censored (e.g., symptom onset may only be reported by date instead of the exact time of day). Second, delays are often right truncated when being estimated in real time (not all events that have occurred have been observed yet). Third, during a rapidly growing or declining outbreak, overrepresentation or underrepresentation, respectively, of recently infected cases in the data can lead to bias in estimates. Studies that estimate delays rarely address all these factors and sometimes report several estimates using different combinations of adjustments, which can lead to conflicting answers and confusion about which estimates are most accurate. In this work, we formulate a checklist of best practices for estimating and reporting epidemiological delays with a focus on the incubation period and serial interval. We also propose strategies for handling common biases and identify areas where more work is needed. Our recommendations can help improve the robustness and utility of reported estimates and provide guidance for the evaluation of estimates for downstream use in transmission models or other analyses.

**Introduction**

Epidemiological delay distributions are key quantities for public health policy [1], clinical practice [2], and infectious disease modeling [3–7]. An epidemiological delay is the time between two epidemiological events. Examples of delays in infectious disease epidemiology include the incubation period (the time between infection and symptom onset), the serial interval (the time between symptom onset in a given infected person and someone they infect), and the generation interval (the time between infection in a given person and someone they infect). Other commonly used epidemiological delays include hospital lengths of stay and delays from symptom onset to hospitalization, hospitalization to death, and symptom onset to specimen date. In the aftermath of the COVID-19 pandemic, there is considerable need for and interest in the estimation of epidemiological parameters, including delays, with at least two R packages [8,9], a World Health Organization working group [10], and meta-analyses of priority pathogens [11–13] working to collect and make these data accessible. In this perspective, we present best practices for the estimation and reporting of epidemiological delays, illustrated by examples for the incubation period and serial interval of directly transmitted (person-to-person) infectious diseases.

The incubation period may vary according to age, vaccination status, underlying conditions [14], and transmission route or intensity of exposure [15,16], among other factors. The serial interval depends on the incubation period, contact patterns in the population, control measures, and the dynamics of infectiousness within a host. While the incubation period must be positive, the serial interval can be positive or negative [17]. The right tail of the incubation period distribution informs the length of quarantine, while the left tail indicates the earliest time symptoms might develop after infection [2]. By comparing the incubation period and serial interval, we can learn about a pathogen's tendency for pre-symptomatic vs. symptomatic transmission [18], which can inform the controllability of an epidemic [19]. At the individual level, pre-symptomatic transmission occurs precisely when the serial interval is shorter than the incubation period of the infectee. In addition, both the incubation period and serial interval are key inputs in mathematical and statistical models, such as those used for nowcasting/forecasting [3,20] or for scenario-based modeling [21,22], which can inform control strategies. The serial interval is often used as a proxy for the generation interval in these applications.

Methods for estimating epidemiological delays have been improving, especially during the COVID-19 pandemic, and recent research has highlighted the importance of appropriately adjusting for three statistical issues inherent in the data collection process: right truncation, censoring, and dynamical bias [23–27]. Indeed, use of delay parameters that are not adjusted for these biases may lead to the propagation of bias into downstream modeling [23] and therefore an incorrect understanding of an epidemic (e.g., over- or under-estimating the risk to the host population) [25].

We provide details about the data needed to estimate delays and how they should be prepared for analysis. Then, we discuss biases that can affect the estimation of epidemiological delay distributions, followed by strategies that can be used to reduce the impact of these biases. Technical details about these biases and how to adjust for them can be found in Park et al. [23].

Finally, we introduce a reporting checklist to aid researchers and reviewers based on our experience estimating epidemiological delay distributions during outbreak responses.

**Data**

Epidemiological delays have start and end times, marked by a primary and secondary event. These events can be observed (e.g., symptom onset time) or unobserved (e.g., usually infection time). Incubation period data include the times of probable exposure and perceived symptom onset for each case. Infection time can be inferred from exposure time because exposure is a necessary, but not sufficient, condition for infection. Data that can be used to estimate the serial interval are those in which symptom onset has been observed for primary and secondary cases. Examples of study designs or public health activities that generate such data include contact tracing [24,28], prospective cohort studies [29–31], household studies [32], or other types of intensive cohort monitoring [33]. Data from passive surveillance, which involves healthcare providers reporting cases to public health agencies, can also be used to estimate delay distributions; however, exposure information may be unreliable because key information may be missing, incomplete, or abstracted from other variables [34].

**Biases in delay data**

Three main biases can affect the estimation of epidemiological delay distributions, 1. censoring, 2. right truncation bias, and 3. dynamical (or epidemic-phase) bias (Table 1).

Censoring is knowing that an event occurred but not precisely when. Data can be right censored (the event is known to have occurred after a certain time), left censored (the event is known to have occurred before a certain time), or interval censored (the event is known to have occurred within a certain time interval). In epidemiological delay data, censoring can affect either primary or secondary events (single interval censoring) or both (double interval censoring) [35]. Epidemiological data are almost always doubly interval-censored due to the time scales of reporting. For example, when reporting occurs daily with a cutoff at midnight, a patient could experience the event of interest (e.g., symptom onset) at any time between 12:00 am and 11:59 pm on a particular day. Moreover, some events are prone to longer censoring intervals than others (e.g., exposure intervals may be longer than one day for cases with multiple possible exposures). Not or incorrectly accounting for censoring of event intervals can lead to biased estimates of a delay [23].

Right truncation is defined as the inability to observe intervals (e.g. incubation periods) greater than some threshold (e.g. greater than the number of days elapsed since infection). It typically applies to real-time settings, when recently occurred events with longer intervals may not have been observed yet, leading to an overrepresentation of shorter intervals when estimating the incubation period or serial interval distribution. However, right truncation should not be confused with right censoring. The latter occurs when we observe the primary event of a case or future case but cannot observe it long enough to witness its secondary event [36], which could be due to, for instance, a study ending prematurely. As a result, we only know that the secondary event did not occur during the observation period and therefore have a right-censored interval for a data point. In contrast, right truncation means that certain intervals are completely missing from our data as observing primary events depends on identifying secondary events first. Right truncation is common in data where case ascertainment depends on the secondary event, e.g. we rarely observe an individual's incubation period until after symptoms develop. Not accounting for right truncation can lead to underestimating the mean delay [23]. Although right truncation is mainly a problem for real-time analyses, retrospective data can be right-truncated if surveillance ended prematurely.

Dynamical bias is another type of common sampling bias which is related to right truncation. During the increasing phase of an epidemic, patients who were infected recently are overrepresented in the recent data, leading to underestimation of delay intervals. Conversely, when the epidemic is decreasing, patients with short delays are underrepresented in the recent data, leading to the overestimation of delay intervals. Dynamical bias is especially problematic during periods of exponential growth and decay of cases when cases are exponentially more and less likely, respectively, to be infected recently rather than further back in time.

**Measuring epidemiological delays**

We aim to estimate the true underlying distribution for each epidemiological delay which characterizes the time between the primary and secondary event. In general, we assume that this distribution does not change over the course of an epidemic (although this may not always be the case [37]). Cases can enter a dataset due to observation of either their primary or secondary event. Regardless of how data were collected, we can organize our data into cohorts using either a forward or backward approach (Figure 1). For the forward approach, we start from primary events that occurred during the same period and prospectively determine when the secondary events occurred—the resulting distribution of the delays is the forward distribution. In contrast, for the backward approach, we start with secondary events that occurred during the same period and retrospectively determine when the primary events occurred—the resulting distribution of the delays is the backward distribution.

Data observed in real-time can be subject to either right truncation or right censoring. Right truncation causes the observed forward distribution to be shorter than the true underlying distribution and has the largest effect when the epidemic is growing because recently infected individuals are overrepresented. In a declining epidemic, right truncation will have a smaller impact on the forward distribution because the proportion of recently infected individuals with unobserved secondary events is lower. Excluding right-censored data from the analysis is equivalent to right truncating the data and leads to underestimation of the delay. Backward distributions are not susceptible to right truncation but can have a delay distribution that is shorter or longer than the forward distribution and the true underlying distribution depending on the phase of the epidemic (i.e. dynamical bias). Both right truncation and dynamical biases are minimal if data from the entire epidemic are available and included in a delay estimate.

Figure 2 shows the impact of different biases on the forward and backward distributions. We recommend always analyzing delay distributions as forward distributions and accounting for potential biases (i.e., censoring and right truncation) as this approach is generally more robust than working with backward distributions and correcting for dynamical bias.

**Adjusting for common biases**

This section is a synthesis of Park et al. [23] in which further details can be found. In general, adjusting for double interval censoring involves estimating the conditional probabilities of the primary and secondary events occurring between their observed lower and upper bounds. Interval censoring should always be adjusted for, and the adjustment method is the same irrespective of the epidemic phase. Right censoring can be handled using methods for survival analysis, such as the Kaplan-Meier approach [36]. Adjusting for right truncation involves normalizing the probability of observing a given delay from the untruncated forward distribution by the probability of observing any delay before the final observation time, and adjusting for dynamical bias involves incorporating the epidemic trajectory (e.g., the growth or decay rate of the epidemic) into the analysis [23,26,38,39].

Although there are several available methods and tools for estimating epidemiological delay distributions, most either have not been validated or do not correct for all potential biases. One example is the *coarseDataTools* R package developed by Reich et al. [35]. This tool has been validated and can correct for double interval censoring, but it does not adjust for right truncation or dynamical bias. In contrast, *epidist*, an R package developed by some of the authors of this study, contains methods which can adjust for all three potential biases. In [23], *epidist* was used to evaluate multiple methods and found that the approximate latent variable censoring and truncation method emerged as the best performer for both real-time and retrospective analyses for most real-world use cases. This method corresponds to the double interval censoring and right truncation adjusted model developed by Ward et al. [24] which we recommend.

Ward et al.'s approach estimates the probability of observing a secondary event conditional on observing the primary event by a given final observation date. Estimated event times for each case are included in the model as unobserved, or latent, variables, and uniform prior distributions are used for both the primary and secondary event times, which can accommodate censoring intervals of arbitrary length. However, when censoring intervals are long, the event time distribution within the censoring interval will deviate from the uniform approximation (as its shape depends on underlying epidemic dynamics) and should be taken into account.

While we recommend Ward et al.'s double interval censoring and right truncation adjusted model for correcting common biases [24], this approach has important limitations; Park et al. found that this method was not able to estimate the mean or standard deviation well in epidemic simulations characterized by very rapid exponential growth and long delays. Details about other possible approaches can be found in Park et al. We suggest avoiding approaches that adjust for right truncation and dynamical biases simultaneously because they lead to overestimation of the mean delay by overcompensating for intervals that have not yet been observed.

**Additional modeling recommendations**

Beyond correctly adjusting for biases, there are several common issues with reported epidemiological delays that may lead to biased conclusions when used in practice or impact their ability to be used at all. Historically, the incubation period and serial interval were often reported using only the mean (and sometimes the range) [40–43]. However, models can be fitted to delay data to adjust for some of the biases we have described and better characterize the tail of the distribution. Assuming a modeling approach is taken, we summarize our recommendations for estimating and reporting for epidemiological delay distributions in Table 2 and give more details in the following two sections.

We recommend fitting a parametric distribution to summarize the empirical delay distribution. Multiple probability distributions should be fitted to delay data [44] and compared using appropriate model comparison criteria (e.g., widely applicable information criterion [WAIC] or leave-one-out information criterion [LOOIC] for Bayesian models). Common distributions for epidemiological delays in the literature include the gamma, lognormal, and Weibull distributions [45]. For delays that can have negative values, distributions that can accept negative values, such as the skew-normal or skew-logistic distributions [46], may be used, or less ideally, the delay data may be shifted to allow for fitting of distributions that only allow positive numbers [47]. Mixture distributions may be appropriate for some delays and should also be considered [48–51].

It is also important to visualize the fitted distributions to check that they fit the data [28]. When doing so, we recommend visualizing the estimated distribution in conjunction with the modeled observation process (e.g., double interval censoring and right truncation). In other words, estimate the latent (continuous) distribution. Then simulate elements of the observation process, such as double interval censoring (e.g. for date-level censoring this will transform continuous delay times into an integer number of days elapsed) and right truncation (this will change the shape of the observed distribution, relative to the latent distribution), from the latent distribution [52]. Not accounting for the observation process after estimating the latent distribution makes visual assessment of the fit difficult, because the observation data and the latent distribution may differ in shape and data type.

Care should be taken when converting the parameters of fitted probability distributions to the summary statistics of interest. For example, the gamma distribution may use either a scale or a rate parameter, in addition to its shape parameter [53], while the standard lognormal parameters, log mean and log sd do *not* correspond to the log of the mean and the log of the standard deviation of the lognormal [54]. Some R packages contain functions to perform parameter conversion, such as EpiNow2 (lognormal) [55], epitrix (gamma) [56], mixR (gamma, lognormal, and Weibull) [57], and epiparameter (gamma, lognormal, Weibull, negative binomial, and geometric) [8].

We recommend stratifying delay estimates whenever there are hypothesized differences across groups as delays such as the incubation period and the serial interval may vary by route of exposure [15], viral species [58] or clade, disease severity, or other factors. Ideally, this stratification should be done jointly in a statistically robust framework [52,59,60]. Wider application of joint modeling approaches could be achieved with more availability of easy-to-use tools [52].

The serial interval and generation interval are two transmission intervals commonly used in infectious disease modeling. For diseases with non-specific initial symptoms, consider estimating delays according to more than one definition of symptom onset (e.g., onset of any symptoms vs rash onset in the case of mpox or measles) [28,61]. Comparing the serial interval and/or generation interval to other transmission intervals (e.g., diagnosis intervals, lab confirmation intervals, or reporting intervals) can provide useful information about the time scales of transmission as long as the interval that is being studied is clearly defined [62].

When Markov chain Monte Carlo methods are used to estimate delay distributions, it is important to visualize posterior predictions against data and check model diagnostics, such as R-hat values, divergent transitions, and effective sample sizes, and report them [63–65]. Convergence issues may indicate that the model is misspecified, making the results unreliable.

**Reporting items**

We recommend reporting an estimate of variability (e.g., standard deviation or dispersion) along with central tendency (e.g., mean or median) for all estimated delay distributions. These quantities are often used as inputs in infectious disease models and can inform both clinical practice and public health policy [2]. The parameter estimates and quantiles of fitted probability distributions should also be reported as they are often used in modeling.

All summary statistics should always be accompanied by credible intervals or confidence intervals for Bayesian and frequentist analyses, respectively (usually 90% or 95% with the width of the reported interval also being reported). High uncertainty in parameter estimates can have substantial impacts on downstream modeling [66,67] and can indicate that more data need to be collected.

Estimates of delays should be accompanied by contextual information to aid in interpretation. For example, we recommend reporting the study sample size; the epidemic curve; which, if any, control measures are in place; and summary statistics on age, sex, geographic location, vaccination status, and possible exposure route(s). The epidemic curve can indicate at which stage of the epidemic the analysis took place and whether the outbreak is now over.

Code and data should be uploaded to repositories, such as GitHub (https://github.com/) or Zenodo (https://zenodo.org/), to ensure reproducibility of the analysis and facilitate re-use of the code. Apart from allowing others to reproduce, validate and potentially improve analyses, providing data along with estimates of delay distributions also ensures that the estimates can be integrated in future pooled estimation efforts as methods continue to be improved. These data should ideally be provided in linelist format, alongside information on censoring and stratifying variables. Importantly, the data should be anonymized to protect patient privacy according to local health data laws and regulations.

If data cannot be shared, we recommend at minimum providing samples of the posterior distribution in a permanent online repository to facilitate future re-analyses (as in [24]). This step is particularly important for distributions with correlated parameters to avoid inflation of uncertainty in downstream use (using parameterizations that have less correlations between parameters, such as those with location and scale, can also help).

**Reporting items specific to the incubation period and serial interval**

In addition to the checklist for reporting epidemiological delay distributions, we recommend additional considerations specific to the incubation period and serial interval (Table 3).

For the incubation period, a case may have had multiple possible exposures prior to symptom onset, especially when community transmission of a pathogen is high. If a case reports multiple exposures, we recommend defining an exposure window that includes all possible exposure dates [1], using disjointed exposure windows where appropriate. Other methods that take this uncertainty into account could be used (such as [68] who used a Bayesian framework to infer the incubation period of the infector). We caution against restricting the analysis to cases with a high degree of certainty about their exposure periods as this can introduce biases [69].

For the serial interval, we only use case pairs where we are fairly confident transmission has occurred [47] (usually based on exposure information collected from patient interviews, see [28,70]). Although this approach could bias the serial interval towards specific lengths of intervals, it is usually preferable to using mis-specified case pairs. In terms of the direction of transmission, a number of approaches can be taken to order the case pairs. Where there is strong evidence that pre-symptomatic transmission does not occur for the disease of interest, reported negative intervals are presumed to be erroneous, and can be removed [28] or reversed [24]. Some studies assume the direction of transmission between epidemiologically linked cases based on the date order of symptom onset [24]; where negative serial intervals are possible, this should be avoided. It is also possible to use genomic data to order the pairs [71,72]. The ideal approach where negative intervals are possible is to use information about pair ordering and fit a distribution that allows for negative values. When no such information is available, a method that does not rely on knowing the order of case pairs would be ideal as this would enable the use of more available data and avoid biases from certain types or durations of exposure being easier or more difficult to link epidemiologically. Although such methods have been developed, they assume a fully sampled population [73,74]. It is important to be transparent about the approach taken. Uncertainty in the source of infection (such as the potential for multiple possible infectors) should also be considered.

**Other considerations**

*How to best use new data.* Delays should be (re-)estimated when possible, especially if current estimates are poor or lacking. New estimates can be compared to those from previous epidemics or from a different phase of the same epidemic. One could consider any new dataset on its own. Alternatively, using mixed effects models to partially pool information across different outbreaks is likely a better use of available data. When choosing an approach, it is important to consider whether the primary modes of transmission or pathogen properties are different than in the past [45,75].

*Choice of prior distributions.* When using Bayesian methods to estimate epidemiological delay distributions, it is important to think about the choice of prior distributions. Some R packages that estimate epidemiological delays, including coarseDataTools [76] and epidist [52], have default prior distributions. Users of these tools should carefully consider the appropriateness of default prior distributions for their analyses and should explore the impact of different prior distributions on their results. We recommend against using uninformed prior distributions, especially uniform prior distributions [77], for the parameters of epidemiological delay distributions. When the delay distribution is already well reported in the literature, it is sensible to use this knowledge to inform prior distributions; however, the methods should be clearly communicated and accompanied by estimates generated from weakly informed prior distributions as sensitivity analyses.

*Meta-analyses*. To reduce uncertainty from small sample sizes, some researchers have combined estimates of epidemiological delays from different studies through meta-analysis [12,13] or pooled analysis (re-analyzing published individual-level data) [2,78]. The latter method is preferred but may not always be possible. If performing a meta-analysis, we recommend performing sensitivity analyses when some estimates from the literature have not been corrected for bias [79]. Published estimates can also be adjusted for bias post-hoc by using the relationship between the backward and forward distributions, as in Park et al. [79]. Some authors have designed custom quality assessment scales to assess bias in studies included in meta-analyses of epidemiological parameters [12,47].

*Time-varying delays*. While some delay distributions are expected to remain constant during an epidemic wave, others can change over time in response to interventions [37] and changes in reporting, among other factors. Specifically, we can study time-varying delays by analyzing changes in forward, rather than backward, delay distributions across cohorts. For example, Ali et al. found that using time-varying estimates of the serial interval of COVID-19 in early 2020 resulted in more accurate estimates of the time-varying reproduction number in mainland China compared to using fixed serial interval distributions [37]. Figure 3 shows a decision tree which can help determine which biases need to be adjusted for depending on the approaches taken for data collection and processing, including whether one is interested in time-varying delays.

**Discussion**

Epidemiological delay distributions are key parameters for preparedness and response to epidemics and pandemics. Their importance has been highlighted during the COVID-19 pandemic [37], the global mpox outbreak in 2022 [1], and other epidemics over the last two decades [7]. We have focused on the incubation period and serial interval, for which estimates are often made at the beginning of an epidemic when contact tracing data are available to support early characterization of the pathogen. While the estimates are most useful for real-time response during this time, they are also the most susceptible to bias. In addition to adjusting for bias, estimates need to be clearly and fully reported to maximize utility and make the most of data that are both costly and difficult to collect [80,81]. Our recommendations can assist with this.

Adjusting for bias when estimating delay distributions is one of the most important recommendations we highlight. Not adjusting for bias can lead to incorrect estimates of delays, which can have direct implications for public health practice. For example, Overton et al. found that the mean incubation period for COVID-19 in early 2020 (corresponding to the ancestral strain of SARS-CoV-2) was 3.49 days without adjusting for right truncation compared to 4.69 days when adjusted [25]. The unadjusted estimate would suggest a quarantine of 10 days would capture 99% of cases, compared to a 14-day quarantine for the adjusted estimate [25]. If the unadjusted estimate from this study had been used to inform the length of quarantine, more still-infectious individuals would have gone on to infect others. Similarly, Park et al. found that ignoring right truncation for a fast-growing epidemic with relatively long delays could result in underestimation of the mean delay distribution by up to 50% [23].

Not adjusting for bias can also lead to incorrect estimates of other parameters of interest that rely on accurate estimates of delays, such as the reproduction number. The basic reproduction number, $R_0$, is the average number of secondary cases caused by a single infected individual in a large, completely susceptible population. If $R_0$ is underestimated, new epidemics may be erroneously perceived as less severe, leading to an insufficient initial public health response.

A limitation of current methods for correcting common biases is that they do not fully account for time-varying changes in delay distributions [23]. Future work on delay distributions or nowcasting (which demonstrates how to model time-varying delays using time-to-event modeling) [82–84] could extend current methods or develop new methods to account for these changes.

To our knowledge, the *epidist* R package [52], developed by some of this paper's authors, is the only software tool that correctly handles interval censoring, right truncation, and dynamical bias. However, it does not currently implement all the best practices we have recommended. A major, though soon to be addressed, limitation is that it can only model lognormal distributions. In addition, while it can handle joint modeling (i.e., across strata) and time-varying delays, these features are not documented and require validation. Its implementation of our recommended method, by Ward et al., does not support custom primary-event prior distributions, so it cannot account for the growth rate in the primary-event window. The tool also scales poorly with increasing amounts of data. Further limitations are discussed in Park et al. [23]. Future development work is planned on the package to address these issues, and community contributions are also welcome.

Many of the best practices outlined in this paper also apply to other epidemiological delays. However, there are issues which we did not cover. For example, we did not focus on methods for estimating delay distributions for vector-borne diseases, as these require additional considerations (e.g., accounting for vector biology) [7,85]. Also, we did not aim to provide a systematic review, and the examples presented in this work were selected to illustrate specific points.

In conclusion, we have provided recommendations for generating and evaluating epidemiological delay distributions. We gave examples of good practice for the incubation period and serial interval from various infectious disease outbreaks over the last few decades; though few examples in the literature incorporate all the best practices outlined in this paper. We hope that our recommendations will provide clarity and structured guidance about what should be reported and how to adjust for biases in delay data. We also hope this checklist will be adopted by the infectious disease modeling community to understand the limitations of existing estimates and improve future estimates.

**Funding statement**

AC is funded by the National Institute for Health and Care Research (NIHR) Health Protection Research Unit in Modelling and Health Economics, a partnership between the UK Health Security Agency, Imperial College London and LSHTM (grant code NIHR200908); and acknowledges funding from the MRC Centre for Global Infectious Disease Analysis (reference MR/X020258/1), funded by the UK Medical Research Council (MRC). This UK funded award is carried out in the frame of the Global Health EDCTP3 Joint Undertaking. SC acknowledges funding support from the Laboratoire d'Excellence Integrative Biology of Emerging Infectious Diseases program (grant ANR-10-LABX-62-IBEID) and the INCEPTION project (PIA/ANR16-CONV-0005). The funders had no role in study design, data collection and analysis, decision to publish, or preparation of the manuscript.

**Disclaimer**

The findings and conclusions in this report are those of the authors and do not necessarily represent the official position of the CDC, U.S. Department of Health and Human Services, NIHR, UK Health Security Agency, or the UK Department of Health and Social Care.

**Author contributions**

Conceptualization: KC, SWP, SA
Data curation: NA
Formal analysis: NA
Funding acquisition: NA
Investigation: KC, SWP, SA
Methodology: NA
Project administration: NA
Resources: NA
Software: NA
Supervision: SA
Validation: NA
Visualization: KC, SWP, SA, AC
Writing - original draft preparation: KC, SWP
Writing - review & editing: All authors

**Data availability statement**

No data were used in this study.

## Tables

Table 1. Adjusting for biases in epidemiological delay distributions. Examples of diseases for each bias were selected based on convenience (either papers written by this study's authors or those encountered during the course of their work). Note that not all methods in this table are discussed in the text, and more details can be found in Park et al. [23].

| Bias | **Interval censoring** | **Right truncation** | **Dynamical bias** |
|---|---|---|---|
| Details | The exact timing of the primary event or secondary event (single interval censoring) or both events (double interval censoring) is unknown (e.g., except for experimental studies, exposure is usually reported as a date or range of dates, rather than a time of day). | Right truncation is a type of sampling bias. It arises because only cases whose secondary event has occurred can be observed. In an ongoing epidemic, right truncation biases the incubation period and serial interval toward shorter intervals because individuals with longer incubation periods may not have developed symptoms or have been reported yet. | Dynamical bias is another type of sampling bias that can be present when case ascertainment is based on the secondary event. It is related to epidemic dynamics: during a growth phase, cases that developed symptoms recently are overrepresented in the observed data, while during a declining phase, these cases are underrepresented. This means that the backward distribution is not representative of the forward distribution during these periods. |
| Impact | Not accounting for interval censoring can lead to biased estimates of a delay's standard deviation. Incorrectly accounting for it can also bias the mean. | Not accounting for right truncation can lead to underestimation of the mean delay [23]. | Not adjusting for dynamical bias when estimating the forward distribution from the backward distribution can lead to under- or over-estimation of delay intervals depending on whether the epidemic is in a growth or declining phase, respectively. |
| Diseases for which this bias has been considered in analyses | Incubation period: mpox [28]; Zika [86]; COVID-19 [87]; 6 vector-borne diseases [78]<br><br>Serial interval: mpox [28] | Incubation period: COVID-19 [25,87]<br><br>Serial interval: mpox [24] | Serial interval: mpox [24] |
| Possible solutions | Use methods, such as Reich et al. 2009 [35], that adjust for double interval censoring; however, this method does not adjust for right truncation. Alternatively, use Ward et al.'s double interval censoring and right truncation adjusted model, which adjusts for interval censoring and right truncation simultaneously [23,24]. | Use an approximate latent variable method, such as Ward et al's double interval censoring and right truncation adjusted model [24] or similar alternatives [23]. | If both primary and secondary event dates are known and the incidence of primary events is changing exponentially at a constant rate, it is possible to use the approach of Verity et al., Britton et al., and Park et al. [26,38,39] to adjust for dynamical bias; however, uncertainty in both growth rate estimates and observed delays need to be taken into account carefully with this approach and the assumption of constant growth rates may not be met in practice. Park et al. present a version of this method that allows for a time-varying growth rate, but it requires untruncated incidence data or assumptions to be made about the recent growth rate [23]. For most settings, considering the forward distribution and accounting for right truncation is recommended. |
| Practices to avoid | Do not adjust for single censoring if the data are doubly interval-censored as this will result in a biased mean [23]. Note that even if both primary and secondary events are reported to have occurred on a single date, the data should be considered doubly interval-censored. | Do not adjust for right truncation and dynamical bias at the same time early in an outbreak as doing this can lead to overadjustment of the downward bias and therefore to overestimation of the delay [23]. | Avoid adjusting for right truncation and dynamical bias at the same time [23]. When analyzing the forward distribution, adjust for right truncation; when analyzing the forward distribution via the backward distribution, adjust for dynamical bias. We recommend the former when possible. |

Table 2. Checklist for reporting/reported epidemiological delay distributions. Examples of diseases for each checklist item were selected based on convenience (either papers written by this study's authors or those encountered during the course of their work).

| Checklist item | Details | Diseases for which this item has been implemented | Possible solutions |
|---|---|---|---|
| **Estimation** | | | |
| Adjust for biases | See Table 1 | See Table 1 | Adjust for censoring (always), right truncation (when needed), and dynamical bias (when needed). Clearly state that all these adjustments have been made, and report both right-truncation-adjusted and right-truncation-unadjusted estimates. Consider using the approach of Reich et al. [35] or Ward et al's double interval censoring corrected model [23,24] to obtain estimates that are not adjusted for right truncation. |
| Compare multiple probability distributions | Estimated delays may depend on the fitted probability distribution, so it is important to use the distribution that best represents the data. | Incubation period: mpox [45]; COVID-19 [87,88]; dengue [85]; HIV [89]; malaria [49]<br><br>Serial interval: MERS [68] | Fit more than one probability distribution to the data [44] and use appropriate model selection criteria to compare them. Visualize the fit of distributions to the data. |
| Correctly convert parameters of probability distributions to summary statistics | Incorrectly converting parameters leads to wrong estimates of delays. | Incubation period: COVID-19 [87] | If writing equations, double check them or use a software package with built-in functions for parameter conversion, such as EpiNow2 (lognormal) [55], epitrix (gamma) [56], mixR (gamma, lognormal, and Weibull) [57], and epiparameter (gamma, lognormal, Weibull, negative binomial, and geometric) [8]. |

| Add subgroups to your model or stratify the estimates whenever appropriate | Estimated delays may vary based on a variety of factors. | Incubation period: Ebola (virus species) [58]; West Nile virus disease (transplant/transfusion vs not) [78]; dengue (serotype) [85]; HIV (sex and age category) [89] | If sample size allows and a difference across groups is hypothesized, add subgroups to your model or stratify the estimates by exposure type, genetic variant/clade, or other factors (e.g., sex, age, vaccination status, etc.). The former approach, also known as joint modeling, is preferred. |
|---|---|---|---|
| Consider estimating other transmission intervals | Comparing other indicators of infections can provide useful information about the time scales of transmission. | Incubation period: mpox (symptom onset definition) [28]; COVID-19 (symptom onset definition) [88]  Serial interval: mpox (symptom onset definition) [28]; measles (symptom onset definition) [61] | Depending on available data, other useful delays may be estimated, such as diagnosis-based, lab-confirmation-based, and report-based transmission intervals. Consider joint modeling approaches for estimation and clearly report the time points used for primary and secondary events. |
| Check model diagnostics | If model requirements are not met, the estimated delay distributions may not be reliable. | Incubation period: dengue [85] | If using Markov chain Monte Carlo methods, make sure all models converge. Visually inspect Markov chain Monte Carlo traces. Check R-hat values, divergent transitions, effective sample sizes, and any other diagnostics and report when appropriate. |
| **Reporting** | | | |
| Report measures of central tendency and variability | These may be used as inputs for infectious disease models and can inform both clinical practice and public health | Incubation period: mpox [45], COVID-19 [87,88]; 6 vector-borne diseases [78]; Zika | Report multiple summary statistics and clearly state which is which. Report at least the mean or median as well as standard deviation, variance, or dispersion. |

| | policy [2]. | [86]; 9 respiratory diseases [2]<br><br>Serial interval: MERS [68]; mpox [28] | |
|---|---|---|---|
| Report quantiles of the probability distribution | The left tail of the incubation period distribution indicates the earliest time symptoms could develop following infection, while the right tail is often of interest for control strategies, such as monitoring people who have been exposed [25,88]. The right tail of the serial interval can also inform the length of quarantine. | Incubation period: mpox [45]; COVID-19 [87,88]; 6 vector-borne diseases [78]; Zika [86]; 9 respiratory diseases [2]<br><br>Serial interval: mpox [24] | Report key quantiles (e.g., 2.5, 5, 25, 50, 75, 95, 97.5, 99) of the distribution in a table. |
| Report the parameters for the fitted probability distributions | These inputs may be used for mathematical modeling and are key for correctly defining the distribution. | Serial interval: mpox [28,90] | E.g. for the gamma distribution, report shape and scale; for the lognormal distribution, report logmean and log standard deviation. If possible, the probability density function should be specified to avoid ambiguity about the parameters. |
| Report uncertainty in the estimates | Communicating uncertainty is a key aspect of outbreak analysis and modeling [67]. Also, high uncertainty in parameter estimates can affect downstream modeling [66]. | Incubation period: mpox [1,45,90]; COVID-19 [87,88]; 6 vector-borne diseases [78]; dengue [85]; Zika [2,7]; MERS [68,91]; 9 respiratory diseases [2]<br><br>Serial interval: mpox [24,28]; COVID-19 [17] | Report 90% or 95% credible intervals or confidence intervals for all estimates (central tendency, variability, quantiles, and parameters for the fitted probability distributions). Ideally, provide joint posterior samples for Bayesian analyses as these are important for characterizing covariance in the posterior distribution. Make sure to report how you have defined your intervals (i.e. report that they are 95% credible intervals). |

| Report characteristics about study sample | Characteristics about the study sample can provide epidemiological context for the estimates which can help with interpretation. | Incubation period: MERS [68]; 9 respiratory diseases [2]  Serial interval: MERS [68] | Report the sample size of the study, demographic characteristics of patients (e.g., age, sex, geographic location, vaccination status), and route of exposure(s) (if known). |
|---|---|---|---|
| Report the epidemic curve and which, if any, control measures are in place | The epidemic curve can provide context about the epidemic phase (increasing, decreasing, or stable) and whether right truncation bias or dynamical bias needs to be considered. | Incubation period: mpox [90]  Serial interval: COVID-19 [38]; MERS [91] | Include a figure of the epidemic curve or provide a reference to the curve on a permanent website (doi). Ideally, the underlying data for the curve would be made available to download. Alternatively, provide an estimate of the growth rate for the study period. |
| Provide anonymized data and documented code | This step improves the reproducibility of the study, and the code can be reused by other teams during future epidemics. It can also facilitate meta-analyses and joint analyses of multiple datasets. | Incubation period: COVID-19 [87,88]  Serial interval: mpox [70] | Deidentified linelist-level data should be provided with relevant stratifying variables. For small epidemics, some authors have reported data relative to an unspecified reference date to protect patient identities [45]. However, the data should still ideally be linked to the epidemic trajectory to address dynamical bias issues. An alternative approach could be to widen the censoring intervals. |

Table 3. Additional checklist items for reporting the incubation period and serial interval.

| Delay | Checklist item | Details | Diseases for which this item has been implemented | Possible solutions |
|---|---|---|---|---|
| Incubation period | Investigate the potential for multiple possible exposures | Sometimes, a case may have had multiple opportunities to be exposed before symptom onset. | Mpox [92]; MERS [68] | For these cases, use an exposure window that includes all possible exposure dates, such as travel to a high-risk area. Use a disjointed exposure window where appropriate (however, this approach may cause some issues for the sampler with Ward et al.'s double interval censoring and right truncation adjusted model [24]). The method developed by Cowling et al. 2015 [68] for MERS could be used as well. |
| Serial interval | Check for negative serial intervals | Negative serial intervals can occur when symptom onset in the infectee occurs before symptom onset in the infector [17]. | Mpox [24,28] | Assuming the data are correct, including negative serial intervals, keep the ordering of the pairs and fit a distribution that allows for negative values (such as normal). If there are negative serial intervals but there is strong evidence that pre-symptomatic transmission does not occur for the disease of interest, consider removing [28] or reversing the order [24] of those case pairs. There are also methods that do not depend on knowing the order of case pairs [73,74]. |
| | Investigate the potential for multiple possible infectors | Sometimes, a case could have been exposed to more than one infected person prior to symptom onset. | Mpox [28,70]; MERS [68] | Restrict the analysis to only cases with a high degree of certainty that the secondary case was infected by the primary case. Then, do a sensitivity analysis with all cases and compare the results. |

**Figures**

Figure 1. Forward and backward approaches for cohorting and analyzing data to estimate epidemiological delay distributions. The y-axis represents unique observations of delays. The yellow circles represent the primary events, while the green squares represent the secondary events. The black horizontal lines represent the delay between primary and secondary events, and the vertical dotted lines show the cohorts. The arrows represent how cases enter the dataset: arrows pointing toward the right indicate that the case's primary event was observed first, while arrows pointing toward the left indicate that the case's secondary event was observed first. Note that the case ascertainment method does not impact the direction we can cohort the data. For forward cohorts (A and C), all primary events that occurred during the same period are selected and prospectively followed until the secondary event occurs. For backward cohorts (B), all secondary events that occurred during the same period are selected; the timing of the primary events is identified retrospectively.

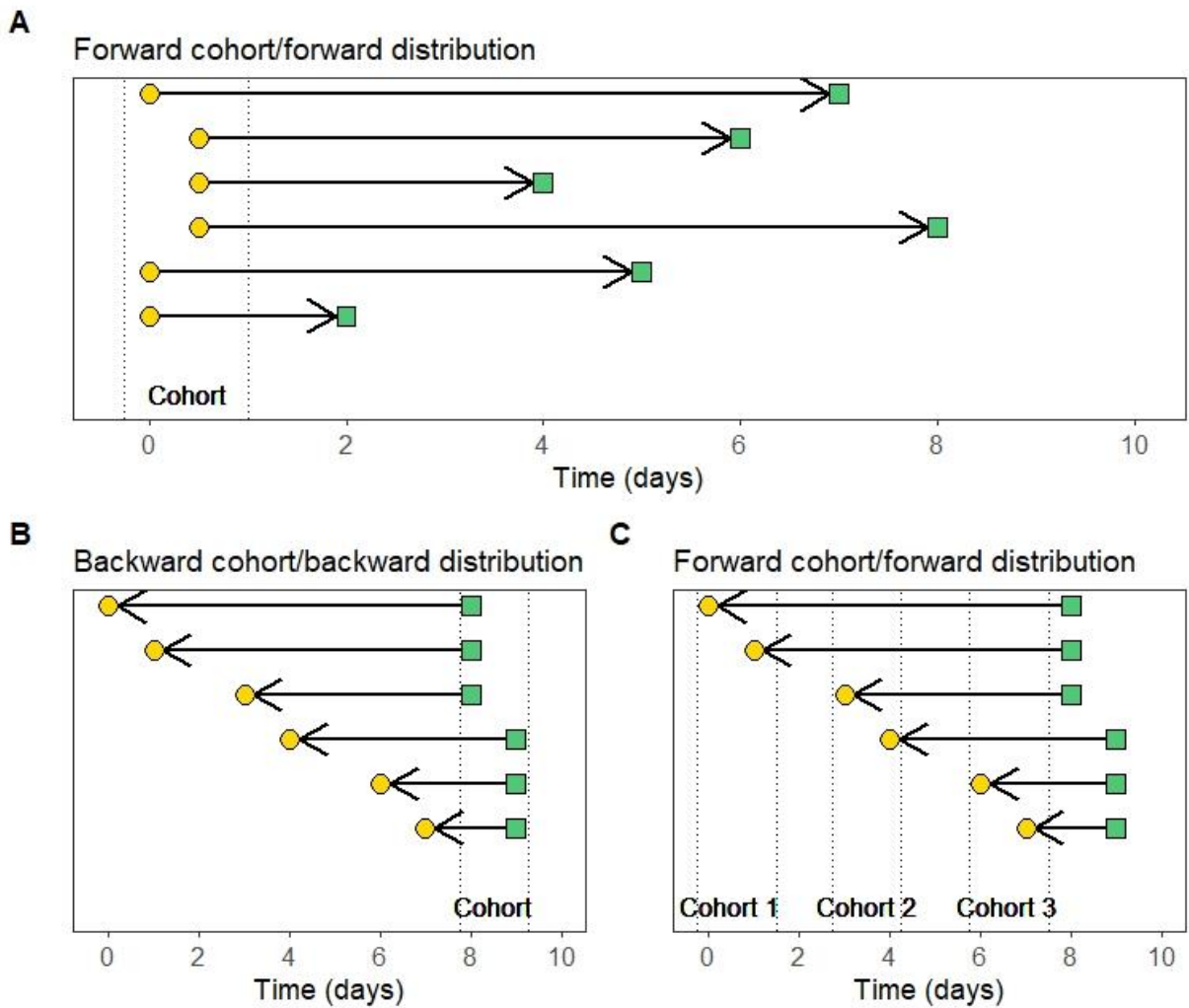

Figure 2. Common biases involved in the estimation of epidemiological delay distributions. The y axis in each panel represents unique observations of delays. The circles represent primary events, while the squares represent secondary events. The horizontal lines represent the delay between events, while the vertical lines represent the time at which the data are observed. The arrows represent how cases enter the dataset: arrows pointing toward the right indicate that the case's primary event was observed first, while arrows pointing toward the left indicate that the case's secondary event was observed first. The brackets "[ ]" represent interval censoring of the primary and secondary events. Delays and events in gray are unobserved. (A) and (B) demonstrate the same scenario, but in (A), observation of the delay is based on the primary event, and there is right censoring, while in (B), observation is based on the secondary event, and there is right truncation. (C) demonstrates an example of a backward distribution in a growing epidemic, when the majority of delays that make up the distribution will be short; hence, the backward distribution will be shorter than the equivalent forward distribution. (D) demonstrates the reverse in a declining epidemic. Both (C) and (D) show the impact of dynamical bias.

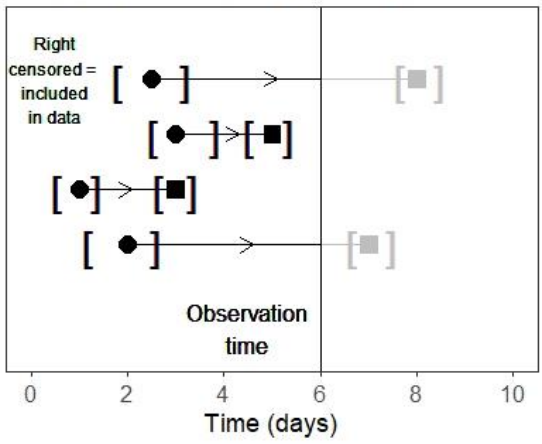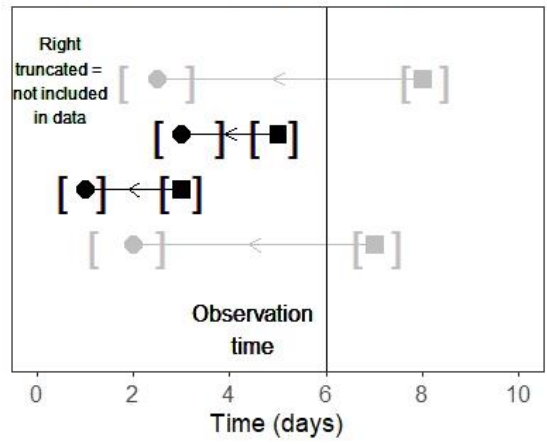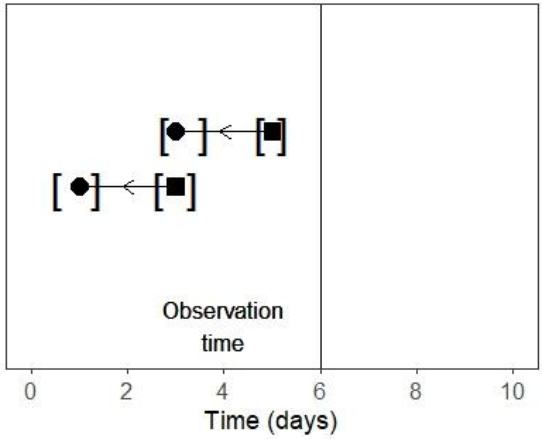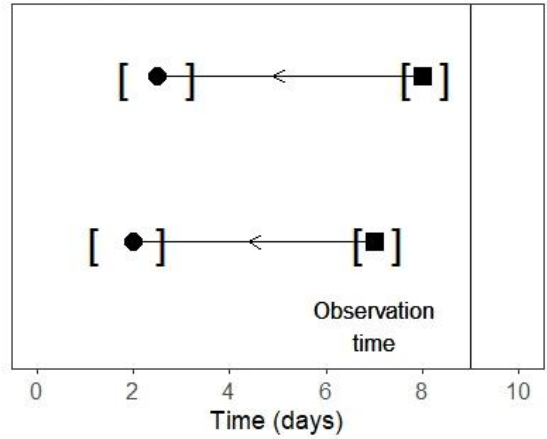

Figure 3. Decision tree for bias adjustment when estimating delay distributions, assuming that double interval censoring is always adjusted for and that the forward distribution is being modeled directly (i.e not via the backwards distribution and dynamical correction) as we recommend regardless of data collection approach. If you have an estimate of the backwards distribution from the literature, see the section on other considerations for advice.

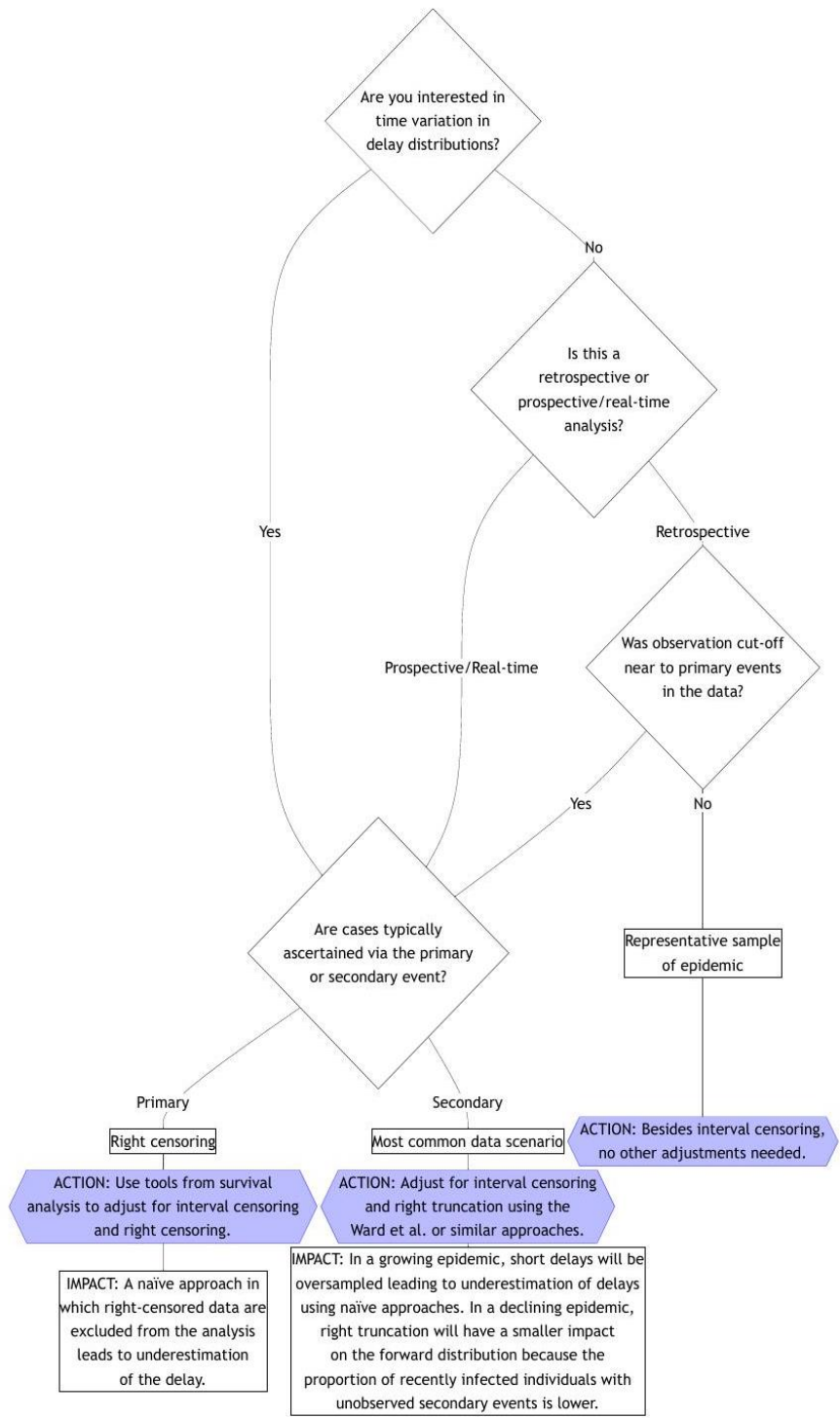